\documentclass[10pt,twocolumn]{article}
\usepackage[T1]{fontenc}
\usepackage[utf8]{inputenc}
\usepackage{cmap,times,microtype,amsmath,amssymb,booktabs,array,graphicx,xcolor}
\usepackage[margin=0.85in,top=0.90in,bottom=0.90in]{geometry}
\usepackage{tikz}
\usetikzlibrary{arrows.meta,positioning,calc}
\usepackage{enumitem,fancyhdr,caption,url}
\usepackage[hidelinks]{hyperref}
\usepackage{natbib}
\newcommand{\PAF}{\textsc{paf}}

\title{\LARGE Beyond the Document:\\ Compiled Context and Governed Agents for Any Work Product}
\author{James Rhodes \qquad George Kang\\AlphaBitCore, Inc.\\\texttt{\{james,george\}@alphabitcore.com}}
\date{November 2026}
\begin{document}
\maketitle
\begin{abstract}
A document preserves the visible result of professional work but often discards the research, calculations, assumptions, evidence, and authorization boundaries behind it. We study a different unit of delivery: a Portable Agent Format (\PAF{}) package that carries a work product together with compiled context, evidence, provenance, and rules for subsequent interaction. To test whether this context remains useful beyond the process that produced it, we evaluate four new packages, created after the hosted and offline readers were fixed: an investment proposal, fund research, financial-statement analysis, and an engineering incident review. Two packages were produced by specialized agents and two through a general-purpose assistant's packaging skill. Across 400 frozen questions and 12,000 evaluated responses, a hosted governed reader succeeds on 96\% of questions versus 86\% with instructions plus raw-file retrieval (Holm-adjusted $p<10^{-7}$). On the 260 answerable questions per condition, success is 94\% versus 79\%; the gap varies by package. Source-material chat and instructions alone reach 48\% and 74\% overall. A separate 0.8-billion-parameter offline reader succeeds on 69\% overall, but 54\% on answerable questions. Its $-1$ response rate is 2.9\%, combining materially wrong and unauthorized answers, versus 56\% with the source and instructions alone. Ablations suggest advantages for the integrated selector and package-adapted writer, without isolating representation from training effects. The same run exposes a failure: offline retrieval resolves none of the 40 historical-record questions in one financial-data package. These findings support treating compiled knowledge as part of the delivered work product, while leaving independent standards interoperability and end-to-end governance conformance unproven.
\end{abstract}

\section{Introduction}
Professional work commonly arrives as a PDF, web page, presentation, or report. The recipient receives a conclusion but not the full state that made it possible: source records, intermediate analysis, alternatives considered, qualifications, evidence lineage, or the rules under which the work was approved. A subsequent question often sends the recipient back to the author or into a general-purpose assistant that can reason about the visible document but does not automatically inherit the producing organization's evidence or authority boundaries. The limitation is not only how questions are answered. It is what gets delivered.

\textbf{Thesis.} The unit of delivery for professional knowledge should be more than a document. A completed work product should travel with its compiled context, evidence and interaction rules, so that another model or runtime can help a recipient interrogate the finished work without reconstructing the original workflow. The producing process and the consuming process need not be the same. Explanation of approved work should also remain distinct from new analysis, new advice or revision requiring fresh authorization.

We study one implementation of this thesis, the Portable Agent Format (\PAF{}). A package combines the presentation the reader sees with typed facts, searchable source records, evidence references, provenance classifications, signed release metadata, governance policy and a runtime profile. Its hosted viewer (Lens) answers against the package under a defined policy; a second reader uses a small local model and a package-specific selection and writing pipeline. The format carries knowledge, not the models or agent processes that produced it. A signature records what was issued; it does not make every statement true.

\textbf{Testable hypothesis.} On questions about a delivered work product, a reader with compiled context and bounded retrieval should recover more of the work's knowledge than a reader with the visible source alone, even after giving that reader the same instructions and ordinary access to the underlying source files. A smaller offline reader should retain useful access to that context, though its retrieval may be less robust.

Earlier exploratory evaluations used packages during reader development, so they could not provide held-out estimates. We therefore created four new packages on 1 November 2026, after fixing the hosted and offline readers, and froze the questions and gold answers before the included evaluation. The packages were not used for reader development or adapter training. We report only these new results, spanning two producer paths in finance and engineering. The conditions and limitations appear in \S\ref{sec:methods} and \S\ref{sec:deviations}.

Four findings organize the paper:
\begin{enumerate}[leftmargin=1.35em,itemsep=2pt]
\item \textbf{Compiled context adds usable knowledge beyond prompting and raw retrieval.} Across four packages, hosted Lens reaches 96\% success against 48\% for ordinary source chat, 74\% for the same source with Lens rules, and 86\% for Lens rules plus BM25 over the source files. The increment over raw retrieval is nonuniform across packages (\S\ref{sec:hosted}).
\item \textbf{The architecture works on different document and production paths.} The evaluation includes two packages produced by Prime agents and two from Claude chats through an ABC-authored packaging skill, including an engineering postmortem. It demonstrates a second path into the same governed publication ecosystem, not independent format conformance (\S\ref{sec:cases}).
\item \textbf{A much smaller reader benefits from packaged context and a purpose-built selector.} On the same packages, the offline 0.8B reader reaches 69\% overall success, including 54\% on answerable questions; 2.9\% of responses receive a $-1$ rating. The ablations implicate retrieval and writer adaptation without fully isolating their effects (\S\ref{sec:offline}).
\item \textbf{Portability of knowledge does not eliminate retrieval failure.} On one package containing numerous near-duplicate XBRL observations, the offline viewer fails all 40 record-level questions despite 100\% hosted record success. This exposes an unresolved selection and disambiguation problem (\S\ref{sec:failure}).
\end{enumerate}

\section{The Portable Work Product}\label{sec:architecture}
\begin{figure*}[t]
\centering
\resizebox{\textwidth}{!}{\begin{tikzpicture}[
  frame/.style={draw,rounded corners=2pt,align=center,fill=black!3,minimum height=2.5em,font=\footnotesize,text width=9.5em},
  arr/.style={-{Latex[length=2mm]},thick}, layer/.style={draw,rounded corners=1.5pt,align=center,fill=blue!5,minimum height=2.1em,font=\scriptsize,text width=8.5em}]
\node[frame] (producer) {Work by agents or people\\\scriptsize external or attested origin};
\node[frame,right=2em of producer] (compiler) {Compile and approve\\\scriptsize facts, records, policies};
\node[frame,right=2em of compiler] (package) {Portable PAF\\\scriptsize signed, versioned package};
\node[frame,right=2em of package] (reader) {Governed reader\\\scriptsize hosted Lens or local viewer};
\draw[arr] (producer)--(compiler);\draw[arr] (compiler)--(package);\draw[arr] (package)--(reader);
\node[layer,below=1.4em of compiler.south,xshift=-3.5em] (pres) {Presentation $P$};
\node[layer,right=.4em of pres] (ctx) {Context $K$};
\node[layer,right=.4em of ctx] (ev) {Evidence $E$};
\node[layer,right=.4em of ev] (gov) {Governance $G$};
\node[layer,right=.4em of gov] (sig) {Manifest / signatures $M,\Sigma$};
\node[layer,right=.4em of sig] (run) {Runtime $R$};
\end{tikzpicture}}
\caption{The delivery unit is the package, not the original conversation or producing agent system. The recipient's reader uses declared context and policy; integrity verification and actual enforcement are separate properties to test.}
\label{fig:system}
\end{figure*}
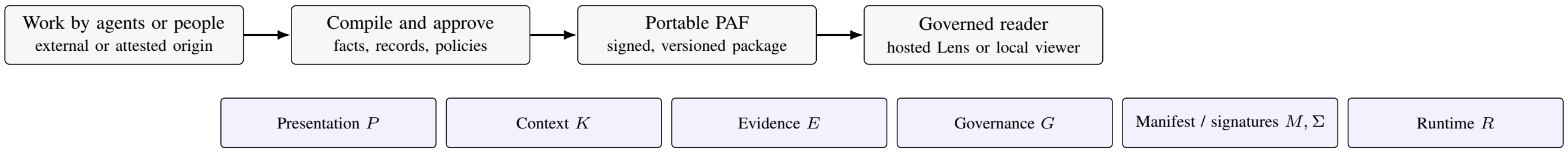

A \PAF{} package $\Pi$ separates six functions. Its \emph{presentation} $P$ is the human-readable deliverable; \emph{compiled context} $K$ contains typed facts, conclusions, rationales and conditions; \emph{evidence} $E$ contains source records, quotations and provenance; \emph{governance} $G$ declares permitted retrieval and interaction; a \emph{manifest} $M$ and signatures $\Sigma$ bind the release to identifiable bytes and signers; and a \emph{runtime profile} $R$ declares how a consumer opens it. The package contains no answering-model weights. These roles are illustrated in Figure~\ref{fig:system}.

\paragraph{Evidence, approval and boundaries.} Each package contains a manifest of payload hashes and both producer and approver signature files. The test bundle provides the four decrypted package archives; their archive SHA-256 digests match the study room records, and the listed payload hashes recompute. This demonstrates consistency of supplied bytes with the manifest, \emph{not} an independent validation of a signature trust chain, approval identity or arbitrary factual accuracy. The packages include provenance records, but an author's declaration of source provenance is not itself verified execution lineage. Content that must not be disclosed should be excluded before issuance: limiting a model's response cannot hide bytes that a recipient can inspect in the package.

\paragraph{Explanation versus revision.} A runtime can use a package to explain information present in the approved context. A question requiring external data, additional computation, a novel recommendation or an altered decision should be declined or routed for new work, not silently rewrite the signed artifact. The evaluation tests the answer-or-decline behavior using a deliberately constructed decline stratum; it does not test end-to-end escalation, revocation, signature-trust enforcement or approval of a successor release.

\section{Four New Packages}\label{sec:cases}
All four packages were generated on 1 November 2026, after the hosted and offline reader implementations had been frozen. Two were produced by an ABC Prime agent workflow; two were constructed in a Claude chat with an ABC-authored \texttt{paf-from-chat} skill and sealed and approved through ABC. These are distinct production paths into the same publication system, not independently specified or independently implemented PAF producers. Table~\ref{tab:packages} summarizes their contents and the comparison documents.

\begin{table*}[t]
\centering\small
\setlength{\tabcolsep}{6pt}
\resizebox{\textwidth}{!}{\begin{tabular}{@{}p{1.25in}p{1.20in}p{2.35in}rrp{1.45in}@{}}
\toprule
Package and subject & Producer path & Baseline supplied to source readers & Typed facts & Records & Presentation in PAF \\
\midrule
Whitaker portfolio proposal & Prime agents & Client proposal PDF & 175 & 2,053 & Proposal PDF, narrative, HTML\\
DFEOX fund analysis & Prime agents & Sponsor-profile HTML reconstruction & 99 & 296 & Fund analysis PDF, narrative, HTML\\
Apple financial statements & Claude chat skill + ABC issue & Two SEC company-concept XBRL JSON files & 19 & 16 & Financial analysis report\\
AWS outage RCA & Claude chat skill + ABC issue & Published AWS post-event summary & 10 & 21 & Engineering postmortem\\
\bottomrule
\end{tabular}}
\caption{Fresh study packages and the input available without the compiled package. The profiles and source exports are files frozen with the study. A source-only baseline is not equivalent to a fully informed author, and the source material differs in complexity across cases. Counts were inspected in the supplied package archives.}
\label{tab:packages}
\end{table*}

\paragraph{Different input and information structures.} The investment proposal contains a multi-account recommendation whose supporting analysis is not fully present in the client PDF. The fund case binds a fund-profile source to separately compiled research. The Apple package starts with two large SEC XBRL company-concept files whose reported periods, fiscal labels, filings and observations contain many near-duplicates; the distilled five-year analysis is far smaller. The engineering package summarizes an AWS incident and its impact windows, starting from an already coherent post-event report. These cases therefore vary in how much compilation or retrieval selection should be required.

\paragraph{Holding the artifacts fixed.} The frozen study lists unique package identifiers and SHA-256 archive digests for all four. Both readers were evaluated on these same published artifacts. The protocol states that the packages had not been used to train, debug or tune the readers. The question sets were written from the packages and frozen before any evaluated answer was generated. This is held-out \emph{package content}, not a random sample of naturally occurring documents, and the questions intentionally cover facts beyond the visible document.

\section{Evaluation Protocol}\label{sec:methods}
\subsection{Questions, scoring and unit of analysis}
Each package has 100 questions: 25 \textsc{memo} questions answerable from its delivered report, 40 \textsc{record} questions whose answer exists in package records but not in that report's headline text, and 35 \textsc{decline} questions requiring an unsupported scenario, new recommendation, outside information or other action the package does not authorize. Candidate memo and record facts were enumerated and drawn using fixed seeds, with logged replacements of unusable candidates; the decline questions were written to fixed topical quotas and audited against the package. A Claude-based question writer and scripted value checks constructed the gold answers. Every model answers each question three times per condition, for 12,000 evaluated answers: 4 packages $\times$ 100 questions $\times$ 3 repeats $\times$ (4 hosted + 6 offline conditions). A further 60 hosted answers per package were regraded as blind anchors in the offline batch; those 240 observations are \emph{not} counted as new study answers.

Two blind judges, Claude Fable 5.1 and Claude Opus 5.5, saw the same anonymized answers with the question, gold, required values and stratum, but not the condition. Answers were scored 2 (correct and complete), 1 (partial), 0 (unhelpful without substantive harm) or $-1$ (wrong or an unauthorized answer to a decline question). Question-level \emph{success} means a score of 2 on at least two of three repetitions. We report the fraction of individual responses scored $-1$ as the \emph{minus-one response rate}, separating materially wrong answers to \textsc{memo}/\textsc{record} questions from unauthorized answers to \textsc{decline} questions. This stratum-based split is derived from judge ratings, not an independent error taxonomy. Neither outcome measures actual financial loss, and a response not scored $-1$ is not necessarily correct or authorized in every respect. The deliberate 25/40/35 weighting is not an estimate of the distribution of real user questions.

\subsection{Hosted comparisons}
All hosted conditions use Claude Sonnet 5.5. They form a ladder of \emph{system} comparisons rather than one-component ablations:
\begin{itemize}[leftmargin=1.15em,itemsep=1pt]
\item \textbf{C1, source chat}: the baseline document(s) in an ordinary assistant, with general knowledge and web access available, ten questions per conversation.
\item \textbf{C2, rules only}: the baseline and Lens's served system instructions, without the package's supporting records or external web access; each question starts a fresh call.
\item \textbf{C4, plain retrieval}: C2 plus BM25 retrieval over the raw files indexed by the package (1,500-character chunks, 200-character overlap, at most 16,000 characters; $k_1=1.5$, $b=0.75$).
\item \textbf{C3, full package}: Lens in the hosted Client Room, over the signed package's compiled context and governed retrieval, one fresh session per question.
\end{itemize}
C3 minus C4 is the practical difference between two complete readers with the same broad underlying source corpus. It does \emph{not} isolate typed-fact compilation from retrieval orchestration, chunking, runtime controls or the use of multiple retrieval steps. C1 to C2 also changes web access and conversation context. We treat C4 as the stronger baseline for incremental value beyond simply making the source files available.

\subsection{Offline comparisons}
The local viewer uses Qwen3.5-0.8B, Q4\_K\_M quantization, llama.cpp, and the frozen v2 selection and writing adapters (sorter v16, writer v7). The six conditions are O1 (source with neutral prompt), O2 (source with Lens instructions), O4 (the ordinary retrieval passages with Lens instructions), O4w (the viewer's trained writer over those raw passages), O5 (the same trained writer over the package's top twelve BM25-ranked lines without its learned selection stage), and O3 (the full viewer with package selection and writing). Sampling is repeated with three fixed seeds. The comparison O5 versus O4w probes the effect of package representation with a common writer; O3 versus O5 probes the added selection stage. Neither is a perfect equal-token-budget ablation, and O3 also includes its integrated response handling.

The two Prime-produced packages declare offline mode disabled, whereas the chat-built packages declare offline mode enabled. The study evaluates answer behavior by opening these fixed artifacts in its offline harness; it does not demonstrate that a conforming viewer would enforce each package's offline-policy flag or verify its signatures against a trust store. Those are separate conformance tests.

\subsection{Statistical comparisons}
Paired success is assessed by exact McNemar tests at question level, with 95\% Newcombe intervals for paired differences. Reported $p$-values are Holm-adjusted within each comparison family: hosted C3--C1, C3--C2, C3--C4; offline O3--O2 and O3--O4w; ablations O5--O4w and O3--O5. We report both individual packages and pooled 400-question results and show the primary Fable judge, with agreement from Opus. Pooling treats questions as the sampled units, not independent package replications; with only four packages, inference to a population of tasks remains limited.

\section{Results}\label{sec:results}
\subsection{Finding 1: compiled context expands reach under a bounded reader}\label{sec:hosted}
\begin{table}[t]
\centering\small\setlength{\tabcolsep}{4pt}
\begin{tabular}{@{}lrrrr@{}}
\toprule
Package & C1 & C2 & C4 & C3 \\
\midrule
Whitaker & 34 & 62 & 80 & \textbf{90}\\
DFEOX & 27 & 51 & 72 & \textbf{96}\\
Apple & 60 & 92 & 97 & \textbf{99}\\
RCA & 71 & 92 & 96 & \textbf{99}\\
\midrule
\textbf{Pooled} & \textbf{48} & \textbf{74} & \textbf{86} & \textbf{96}\\
\bottomrule
\end{tabular}
\caption{Hosted success (\%, score 2 in at least two of three repetitions). Primary blind judge; 100 questions per package. Pooled displayed rates are rounded to whole percent.}
\label{tab:hosted}
\end{table}

Across the four packages, the hosted full-package condition succeeds on 96\% of questions, compared with 48\% for ordinary source chat, 74.25\% with Lens instructions alone, and 86.25\% when raw-file retrieval is added. Table~\ref{tab:hosted} shows the large range across cases. The paired pooled differences are $+48$ percentage points versus C1 (95\% CI 43--53, Holm $p=2.5\times10^{-52}$), $+21.75$ versus C2 (CI 18--26, $p=1.7\times10^{-21}$), and $+9.75$ versus C4 (CI 7--13, $p=1.8\times10^{-8}$). The second judge gives the same directional findings.

The result against plain retrieval is not uniform. On Whitaker, C3 exceeds C4 by 10 points (Holm $p=0.013$), and on DFEOX by 24 ($p=3.0\times10^{-6}$). On Apple it adds 2 ($p=0.62$), and on RCA 3 ($p=0.38$). Thus the pooled difference does not establish a similarly sized incremental benefit in every kind of package. The largest raw-document gaps occur where the source artifact withholds much of the finished research; the smallest occur where the source files already offer structured facts or a well-written post-event explanation.

The 35\% \textsc{decline} quota rewards appropriate refusals. On the 260 answerable questions per condition (\textsc{memo} plus \textsc{record}), C3 reaches 245/260 (94.2\%) against C4's 206/260 (79.2\%), a descriptive 15.0-point difference. C1 reaches 64.6\% and C2 61.2\% on answerable questions despite C2's higher overall rate: instructions improve refusal behavior without adding missing facts. Within the record-only stratum, C3 reaches 92\%, C4 77\% and C2 56\%. Those questions intentionally sample facts beyond the visible report, so C4, which receives raw-source retrieval, is the more informative control.

\begin{table}[t]
\centering\small\setlength{\tabcolsep}{5pt}
\begin{tabular}{@{}lrrrr@{}}
\toprule
Measure & C1 & C2 & C4 & C3\\
\midrule
MEMO success (\%) &72&70&83&98\\
RECORD success (\%) &60&56&77&92\\
Answerable success (\%) &65&61&79&94\\
DECLINE success (\%) &17&99&99&99\\
$-1$: wrong answer ($n$) &35&19&5&1\\
$-1$: unauthorized answer ($n$) &205&3&2&2\\
Total $-1$ responses (\%) &20.0&1.83&0.58&0.25\\
\bottomrule
\end{tabular}
\caption{Hosted pooled outcomes. Answerable combines 25 \textsc{memo} and 40 \textsc{record} questions per package (260 pooled). Wrong-answer and unauthorized-answer counts partition $-1$ ratings by answerable versus decline stratum; each condition has 1,200 responses.}
\label{tab:strata}
\end{table}

\paragraph{Correct answers versus authorized refusals.} Of C1's 240 responses scored $-1$, 205 answer questions that should have been declined and 35 are materially wrong answers to answerable questions. C3 has three such responses in total: one wrong answer and two unauthorized answers. Most of the reduction therefore reflects better adherence to the intended boundary, not merely improved factual accuracy. Rules already account for most of the gain on declines, while compiled context and retrieval increase answerable-question success. The experiment does not isolate deterministic gating from prompts or retrieval, and these ratings do not establish universal policy conformance.

\subsection{Finding 2: one package contract spans production paths}\label{sec:producer}
Both hosted Lens and the offline reader consumed packages produced from ABC Prime workflows and packages authored in a Claude chat then sealed by ABC. The Apple and RCA hosted conditions each reach 99\% success; C4 is already at 97\% and 96\%, respectively. These packages establish cross-\emph{path} consumption, including a case outside financial services, without demonstrating fully independent production or cross-vendor conformance. The packaging skill, the signing pipeline, the format schemas and both readers remain within the ABC ecosystem.

A key architectural observation is distinct from a performance comparison: the consuming reader does not rerun the producing workflow. It uses the delivered package's facts, records, evidence and rules. The study does not directly measure recipient adoption, operating cost, customer trust, cryptographic trust-chain enforcement or the quality of the original producing work; those require separate tests.

\subsection{Finding 3: a small offline model retains much of the packaged knowledge}\label{sec:offline}
\begin{table}[t]
\centering\small\setlength{\tabcolsep}{4pt}
\begin{tabular}{@{}lrrrrrr@{}}
\toprule
Package & O1 & O2 & O4 & O4w & O5 & O3\\
\midrule
Whitaker &4&13&5&36&48&\textbf{64}\\
DFEOX &11&24&17&35&66&\textbf{77}\\
Apple &3&7&22&37&38&\textbf{51}\\
RCA &17&37&28&37&52&\textbf{84}\\
\midrule
\textbf{Pooled} &\textbf{9}&\textbf{20}&\textbf{18}&\textbf{36}&\textbf{51}&\textbf{69}\\
Answerable only &12&20&23&5&28&54\\
\bottomrule
\end{tabular}
\caption{Offline success (\%) on 400 questions; the final row restricts to 260 answerable questions. O4w and O5 use the frozen trained writer; O3 also uses the package selector.}
\label{tab:offline}
\end{table}

The integrated 0.8B offline viewer (O3) succeeds on 69\% of all questions, compared with 20.25\% for the same small model given the baseline and Lens instructions (O2), and 36.25\% for the trained viewer writer given raw-file passages (O4w). The paired differences are $+48.75$ points (95\% CI 43--54, Holm $p=7.3\times10^{-45}$) and $+32.75$ (CI 28--37, $p=3.2\times10^{-34}$), respectively. On answerable questions alone, O3 reaches 141/260 (54.2\%), O2 reaches 51/260 (19.6\%), and O4w only 13/260 (5.0\%). Of O3's 276 overall successes, 135 are correct declines, so the 69\% pooled rate should not be read as 69\% answer coverage. Responses scored $-1$ fall from 56.33\% under O2 to 2.92\% under O3; these include both wrong answers and unauthorized responses, separated in Table~\ref{tab:offstrata}. These are within-model comparisons, not a claim of parity with the hosted system.

The offline ablation helps locate the performance difference, but is not a clean test of representation alone. The same trained writer increases from 36.25\% over raw BM25 passages (O4w) to 51\% over the package's BM25 lines (O5), a $+14.75$-point difference (CI 10--19, Holm $p=4.4\times10^{-10}$). The trained writer obtains only 8\% \textsc{memo} success with raw passages, below the unadapted O2 reader's 20\%. A plausible explanation is distribution mismatch: the writer was adapted to package-style input, not arbitrary raw-file chunks, but this was not independently tested. Restoring the viewer's learned selection stage (O3) adds another 18 points over O5 (CI 14--22, Holm $p=3.7\times10^{-14}$). These comparisons demonstrate differences between integrated configurations; the observed gains may reflect representation, training compatibility, retrieval selection and context budgets together.

\begin{table}[t]
\centering\small\setlength{\tabcolsep}{4pt}
\begin{tabular}{@{}lrrrr@{}}
\toprule
Outcome & O2 & O4w & O5 & O3\\
\midrule
MEMO success (\%) &20&8&54&80\\
RECORD success (\%) &19&3&13&38\\
Answerable success (\%) &20&5&28&54\\
DECLINE success (\%) &21&94&93&96\\
$-1$: wrong answer ($n$) &469&64&74&27\\
$-1$: unauthorized answer ($n$) &207&10&10&8\\
Total $-1$ responses (\%) &56.3&6.2&7.0&2.9\\
\bottomrule
\end{tabular}
\caption{Offline pooled outcomes. Answerable success uses 260 \textsc{memo} and \textsc{record} questions. Wrong-answer and unauthorized-answer counts partition $-1$ ratings by question stratum (1,200 responses per condition). Decline success materially boosts pooled rates, especially O4w.}
\label{tab:offstrata}
\end{table}

\subsection{Finding 4: the held-out run reveals a retrieval failure}\label{sec:failure}
In the Apple package, the offline viewer's overall success is 51\%, but its success on the 40 \textsc{record} questions is \emph{0\%}. On 114 of the 120 individual record-answer attempts it responds: ``The proposal doesn't state that. I can pass the question to the sender.'' By contrast, hosted Lens succeeds on 100\% of those 40 questions. The questions ask for specific reporting periods and filings among many near-duplicate observations of revenue and net income. The frozen offline selector does not reliably map the reader's date and period qualifiers to the precise XBRL row; it can find nearby amounts but not the requested fact. A conservative writer therefore often declines instead of hallucinating, at a large cost in coverage.

This failure also explains why the Apple ablation is different: O5 and O4w differ by only one overall point, with no statistically detectable advantage for package lines alone. The problem was not fixed before collecting the results. Altering the selector after observing it would have converted this set into another development evaluation. The result demonstrates both the utility and the limits of an offline, compact reader: a packaged body of evidence can be portable while precise record selection is not automatically portable across runtimes.

\subsection{Judge stability and audit checks}
Both judges see the same blind items; pooled success differs modestly (hosted C3 96\% versus 95.5\%; offline O3 69\% versus 67\%), with no main pooled comparison changing direction or statistical conclusion. Pass/fail agreement is 98.5--99.2\% by package for the hosted batches and 95.3--98.0\% for offline batches including anchors; Cohen's $\kappa$ spans 0.86--0.95. A 60-answer hosted anchor per package was blindly regraded with the offline batches, with 97--100\% within-judge pass/fail matching. This addresses grader drift across batches but does not establish independent human validation: both judges and the question writer are from the Claude model family.

We independently recomputed success and harm summaries from all eight graded CSV files (4,800 hosted and 7,200 offline responses), checking the three-repeat design, conditions, scores and package-archive digests. The published summary rates reconcile to the underlying rows within rounding. We did not independently rerun the hosted models or validate the underlying source claims against external primary documents.

\section{Threats to Validity and Scope}\label{sec:deviations}
\paragraph{Held-out evaluation.} The four packages were created after the reader implementations and offline adapters were fixed and were not used in earlier reader training or debugging. Each package was evaluated using 100 preselected questions with frozen gold answers, three fixed repetitions per condition, the hosted Lens instructions applicable to the study, and blinded grading by two judges. Hosted and offline answers were graded in separate blind batches, with hosted anchor responses repeated in the offline batch to check grading stability. The included results were generated with the study conditions described in \S\ref{sec:methods}; the held-out claim concerns the package content and fixed model configurations, not independence of package authorship or judging.

\paragraph{Question construction and sampling.} The golds were compiled from the PAFs and their supporting evidence by a Claude question writer. The result therefore measures accurate access to a package-defined body of knowledge, not independent truth. The record-only sampling intentionally selects data omitted from the visible report and can favor a package-aware reader; a strong retriever over identical raw sources is the essential control. The fixed 35\% decline quota can inflate the success of conservative readers relative to typical question traffic, while majority-of-three scoring can mask intermittent wrong answers. The same underlying case facts are repeated across three runs and should not be treated as independent question units.

\paragraph{Systems are not component-matched.} The hosted reader can perform governed, structured or multi-stage retrieval while C4 uses a single BM25 pass; the 9.75-point increment therefore reflects the \emph{complete system}, not just the file format or a deterministic claim checker. O4w and O5 share a writer but not matched input distributions: the adapted writer may handle structured package lines better than raw passages because of its training. Segmentation and context budgets also differ; the offline system uses trained adapters and an embedding-based selector, and the study has no base-model-on-PAF condition that isolates those adapters. Apple O4w also received a longer context slot solely to avoid truncating an oversized raw-file prompt. C1 differs from C2 in web access and conversation batching as well as instructions. The four packages and one answering provider limit cross-task and cross-model generalization.

\paragraph{Security and governance are proposed properties, not validated outcomes.} We checked digest consistency for the supplied PAF archives and manifest entries, not an independent Ed25519 trust chain or the authority of the human approver. Some study packages declare offline mode disabled, and the test harness does not establish a conforming offline verifier. This study does not test tamper refusal, revocation, malicious package contents, prompt injection, audience-safe byte projection, escalation into approved revision, or formal execution attestation. A low $-1$ answer rate is evidence of behavior on these questions; it does not prove cryptographic correctness, soundness of arbitrary material claims, or suitability of financial advice.

\paragraph{Evidence availability.} The authors retain the frozen protocols, question sets and golds, individual answers, both judges' blind ratings and explanations, results tables, scripts and PAF packages in a private research bundle. These files are not publicly released. This limits independent replication of both the experimental findings and the provenance assertions.

\section{Interpretation: what travels, and what fails to travel}
\subsection{The gain is not merely a stricter prompt}
The comparison ladder distinguishes instruction-following from access to compiled knowledge. On the hosted system, the shift from ordinary chat (C1) to Lens instructions without the added records (C2) improves out-of-scope declines from 17\% to 99\%, while record success remains approximately unchanged or declines slightly (60\% to 56\%). This is consistent with a prompt that narrows unauthorized assertions without supplying the additional information the recipient needs. Adding ordinary retrieval (C4) takes record success to 77\%; the bound package (C3) reaches 92\%. Because the corpus is largely the same, this last comparison is evidence that \emph{how} the knowledge is organized and made reachable matters, but the experiment does not uniquely attribute the difference to typed facts rather than the hosted runtime's retrieval algorithm.

On the offline model the picture is more extreme. O2 produces 20\% overall success and 56\% responses rated $-1$, including 469 wrong answers and 207 unauthorized answers. O4w lowers the $-1$ rate to 6.2\% largely by refusing questions, but resolves only 5\% of answerable questions; O5 resolves 28\%, and O3 54\%. These results support evaluating \emph{answer coverage} separately from refusal policy. They suggest that learned selection and training-compatible representations matter to the integrated reader; the test does not measure their independent causal contributions.

\subsection{Human-authored and AI-authored documents}
The experiment deliberately avoids limiting PAF to the native outputs of the Prime agents. The DFEOX case compiles research around a source profile that is not the engagement's own memo; the Apple and RCA cases compile content prepared in a general-purpose assistant, with ABC acting as packaging and issuance service. The visible work product is therefore not required to have been authored by the package's eventual consumer, and a recipient does not need the original agent environment to ask a question. The study does not show that a source page was edited in place, nor does it establish all possible document types. It shows the narrower, consequential property that separate original content and compiled context can be delivered as one usable artifact.

This is relevant to organizations that already have workflows producing PDF reports, research notes, audit records or incident summaries. A transition to interactive, bounded work products need not start by replacing every authoring system. It could start with an issuing process that attaches compiled context and declared authority to what existing processes already produce. Whether that lowers costs or improves service quality requires deployment studies beyond this question-answering benchmark.

\subsection{A low observed harm rate is not a semantic proof}
The $-1$ response measure combines two different behaviors: materially wrong answers on answerable questions and unauthorized answers on questions that require refusal. We therefore separate these counts, rather than describing their sum as a factual-error or real-world harm rate. The ratings do not audit the runtime's complete call trace, prove that every retrieved claim derives from a permissible source, or check who was entitled to see each packaged byte. Likewise, a verifier can establish that a signed manifest identifies certain bytes without establishing that the numbers in them are true. The distinction is especially important in regulated use: \emph{integrity}, \emph{provenance}, \emph{authorization}, \emph{answer grounding}, and \emph{substantive correctness} are different assertions and require different tests.

\subsection{Next tests suggested by the held-out failure}
The Apple case indicates a concrete next step: a held-out experiment that tests selectors over repeated numeric observations with explicit keys for issuer, metric, filing, fiscal period, start and end dates, units and amendments. Such a selector should retrieve the correct row or visibly return no confident match, with a held-out test set used only after the new algorithm is fixed. Separately, PAF verification needs mutation tests (tampered payloads, mismatched signatures, revoked issuers, invalid approvals) and a viewer-conformance suite covering offline policy and audience projection. Those studies can validate whether the format's broader trust claim survives adversarial and independent implementations. They should not be read back into the present evaluation as already proven.

\section{Related Work}
\paragraph{Document conversation and retrieval.} Retrieval-augmented generation (RAG) conditions a model on source passages \citep{lewis2020rag}. The strongest baseline here is explicitly a prompted reader with BM25 over the package's raw source files. The measured contribution beyond that baseline must be interpreted as the joint effect of compiled context, retrieval architecture and runtime behavior rather than evidence of a universally superior retriever. Atomic factuality measurement \citep{min2023factscore} and approaches that check numeric claims against a structured evidence set \citep{pcn2025} complement, but do not substitute for, delivery of a persistent compiled artifact.

\paragraph{Integrity, provenance and authority.} C2PA content credentials and W3C Verifiable Credentials bind statements and identifiers to signed artifacts \citep{c2pa2025,w3cvc2025}; in-toto and SLSA establish software build provenance \citep{torresarias2019intoto,slsa2023}. Those mechanisms provide important building blocks but do not alone define the context available to a later answering agent. The \PAF{} architecture combines content, evidence references, release metadata and declared interaction policy in one delivery unit. A signature authenticates the signed statement, not the correctness of all underlying facts. Prompt injection against retrieved evidence remains a threat \citep{greshake2023injection}.

\paragraph{Agent execution and compiled memory.} Proof of Execution \citep{rhodes2026poe} frames signed runtime evidence for governed actions, while compiled memory \citep{rhodes2026atlas} treats verified experience as reusable context for agents. The present paper concerns a different boundary: transferring useful knowledge \emph{with the work product} so that the recipient's reader need not reconstruct the original producer. Policy-bound answer receipts \citep{leray2025pgrag} and approval-integrity mechanisms \citep{approval2026} are complementary. Neither independent execution replay nor a cross-organization trust registry is evaluated in this study.

\section{Toward an Interoperable Format}
The four-case test supports the value of a package carrying context across two producer paths and two readers, but all implementations and the institutional signing path share ABC-controlled format components. A generally interoperable format would require a public schema, canonical serialization, well-defined provenance semantics, trust roots and revocation rules, explicit offline permissions, audience-specific evidence projection, and independent verifier and runtime conformance tests. In particular, conformity cannot be inferred from a reader's ability to parse and answer questions about a package. Trusted issuance must separate \emph{who signed}, \emph{what was approved}, \emph{which evidence was checked}, and \emph{what the consuming runtime actually enforced}.

The offline Apple failure identifies a concrete semantic-interoperability requirement that goes beyond valid bytes: record selectors must preserve period, filing, unit, instrument and qualifier identity. Better labeling, typed lookup or an explicit structured query interface may help, but these were not evaluated in this run. The next experiment should repeat the exact question set under independent implementations and separately test cryptographic and policy conformance, without tuning on the test packages.

\section{Conclusion}
A finished document need not be the final container for professional knowledge. The clean evaluation of four newly produced packages demonstrates that compiled context can travel with a deliverable and remain useful to a hosted governed reader and to a much smaller offline model. The hosted system resolves 96\% of the fixed question set (94\% on answerable questions), compared with 86\% (79\% answerable) when the same model receives raw-source retrieval. The local 0.8B reader resolves 69\% overall (54\% answerable), compared with 36\% overall (5\% answerable) for its trained writer over raw passages. These are results for the studied systems and question distribution, not guarantees for other tasks.

The ambition is not to make every question answerable. It is to preserve the knowledge and authority boundaries behind completed work so that readers can request explanations from an accountable body of evidence and recognize when new work or a new approval is required. The held-out run also shows why that distinction matters: when the offline reader could not identify the correct XBRL record, it mostly declined rather than inventing a figure, sacrificing coverage. Portability of the artifact is necessary but not sufficient; retrieval and governance need independent conformance. A document records what was decided. A PAF package can carry the knowledge behind that decision into the next interaction.

\paragraph{Data and code availability.} The study records and implementation materials are retained privately by the authors and are not publicly released. The manuscript reports aggregate outcomes and the protocol details needed to interpret the results but does not make the research bundle available for independent replication.

{\footnotesize\bibliographystyle{plainnat}\bibliography{refs}}

@misc{approval2026,author={Alpay, Faruk and Alpay, Taylan},title={Approval integrity and recovery in LLM answer publication},year={2026},note={arXiv:2609.15576}}

@misc{c2pa2025,author={{Coalition for Content Provenance and Authenticity}},title={Content credentials: C2PA technical specification, version 2.2},year={2025},month={May},howpublished={\url{https://spec.c2pa.org/specifications/specifications/2.2/specs/C2PA_Specification.html}}}

@inproceedings{greshake2023injection,author={Greshake, Kai and Abdelnabi, Sahar and Mishra, Shailesh and Endres, Christoph and Holz, Thorsten and Fritz, Mario},title={Not what you've signed up for: Compromising real-world LLM-integrated applications with indirect prompt injection},booktitle={Proceedings of the 16th ACM Workshop on Artificial Intelligence and Security (AISec)},year={2023},note={arXiv:2302.12173}}

@misc{leray2025pgrag,author={Le Ray, Jean-Marie},title={Policy-governed RAG: Research design study},note={arXiv:2510.19877 [cs.CR]},year={2025}}

@inproceedings{lewis2020rag,author={Lewis, Patrick and Perez, Ethan and Piktus, Aleksandra and Petroni, Fabio and Karpukhin, Vladimir and Goyal, Naman and Kuttler, Heinrich and Lewis, Mike and Yih, Wen-tau and Rocktaschel, Tim and Riedel, Sebastian and Kiela, Douwe},title={Retrieval-augmented generation for knowledge-intensive NLP tasks},booktitle={Advances in Neural Information Processing Systems 33},year={2020},note={arXiv:2005.11401}}

@inproceedings{min2023factscore,author={Min, Sewon and Krishna, Kalpesh and Lyu, Xinxi and Lewis, Mike and Yih, Wen-tau and Koh, Pang Wei and Iyyer, Mohit and Zettlemoyer, Luke and Hajishirzi, Hannaneh},title={FActScore: Fine-grained atomic evaluation of factual precision in long form text generation},booktitle={Proceedings of the 2023 Conference on Empirical Methods in Natural Language Processing},year={2023},note={arXiv:2305.14251}}

@misc{slsa2023,author={{Open Source Security Foundation}},title={Supply-chain levels for software artifacts (SLSA) specification, version 1.0},howpublished={\url{https://slsa.dev/spec/v1.0/}},year={2023}}

@misc{rhodes2026atlas,author={Rhodes, James and Kang, George},title={Compiled memory: Not more information, but more precise instructions for language agents},note={arXiv:2603.15666 [cs.AI]},year={2026}}

@misc{rhodes2026poe,author={Rhodes, James and Kang, George},title={Proof of execution: Runtime verification for governed AI agent actions},note={arXiv:2607.05397 [cs.CR]},year={2026}}

@misc{pcn2025,author={Solatorio, Aivin V.},title={Proof-carrying numbers (PCN): A protocol for trustworthy numeric answers from LLMs via claim verification},note={arXiv:2509.06902},year={2025}}

@inproceedings{torresarias2019intoto,author={Torres-Arias, Santiago and Afzali, Hammad and Kuppusamy, Trishank Karthik and Curtmola, Reza and Cappos, Justin},title={in-toto: Providing farm-to-table guarantees for bits and bytes},booktitle={28th USENIX Security Symposium},pages={1393--1410},year={2019}}

@misc{w3cvc2025,author={{World Wide Web Consortium}},title={Verifiable credentials data model v2.0},note={W3C Recommendation, May 2025},howpublished={\url{https://www.w3.org/TR/vc-data-model-2.0/}},year={2025}}
\end{document}